\documentclass[aps,pre,onecolumn,reprint,
]{revtex4-2}
\usepackage[utf8]{inputenc}
\usepackage{bm}
\usepackage{amssymb}
\usepackage{mathtools}
\usepackage{amsmath}
\usepackage{xcolor}
\usepackage{enumitem}
\usepackage{natbib}
\usepackage{hyperref}
\usepackage{graphicx}
\usepackage{ragged2e} 

 \hypersetup{colorlinks=true, linkcolor=blue, breaklinks=true, urlcolor=blue, citecolor=blue}
\usepackage{lipsum}
\usepackage[ruled,vlined]{algorithm2e}

\def \bu {\pmb{u}}
\def \bU {\pmb{U}}
\def \bx {\pmb{x}}
\def \by {\pmb{y}}

\def \Tr {\text{Tr}}
\usepackage{rotating}
\usepackage[english]{babel}
\usepackage{amsthm}

\theoremstyle{definition}

\newtheorem{remark}{Remark}

\usepackage{subfig}

\makeatletter
\renewcommand{\fnum@figure}{FIG. \thefigure}
\makeatother

\makeatletter
\providecommand{\@LN}[2]{}
\makeatother

\makeatletter
\long\def\@makecaption#1#2{%
  \vskip\abovecaptionskip
  \sbox\@tempboxa{#1: #2}%
  \ifdim \wd\@tempboxa >\hsize
    {\justifying #1: #2\par}
  \else
    \global\@minipagefalse
    \hb@xt@\hsize{\justifying\hfil #1: #2\hfil}%
  \fi
  \vskip\belowcaptionskip}
\makeatother

\usepackage{comment}
\usepackage[normalem]{ulem}  

\newcommand{\redstrike}[1]{%
  \bgroup
  \markoverwith{{\rule[0.6ex]{2pt}{0.8pt}}}%
  \ULon{#1}%
  \egroup
}

\usepackage{multirow}

\begin{document}

\title{
\textbf{Open Networks in Discrete Time: Passing vs Blocking Behavior}
}

\author{Amirhossein Nazerian$^{1}$, Malbor Asllani$^{2}$, Melvyn Tyloo$^{3,4}$, Francesco Sorrentino$^{1,*}$  \\
\normalsize $^{1}$Department of Mechanical Engineering, University of New Mexico, Albuquerque, NM 87131, United States of America\\
\normalsize $^{2}$Department of Mathematics, Florida State University, 1017 Academic Way, Tallahassee, FL 32306, United States of America \\
\normalsize $^{3}$
Living Systems Institute, University of Exeter, Exeter, EX4 4QD, United Kingdom \\
\normalsize $^{4}$
Department of Mathematics and Statistics, Faculty of Environment, Science, and Economy, University of Exeter, Exeter, EX4 4QD, United Kingdom\\
\normalsize $^{*}$ e-mail: fsorrent@unm.edu}



\begin{abstract}
   This paper presents a unified framework for analyzing the input-output behavior of discrete-time complex networks viewed as open systems. { Importantly, we focus on systems that are inherently modeled in discrete time—such as opinion dynamics, Markov chains, diffusion on networks, and population models—reflecting their natural formulation in many real-world contexts.} By an open network we mean one that is coupled to its environment, through both external signals that are received by designated input nodes and response signals that are released back into the environment via a separate set of output nodes.
   We develop a general framework for characterizing whether such networks amplify (pass) or suppress (block) the external inputs. Our approach combines the transfer function of the network with the discrete-time controllability Gramian, using the $\mathcal{H}\textsubscript{2}$-norm to quantify signal amplification. We introduce a computationally efficient network index based on the Gramian trace and eigenvalues, enabling scalable comparisons across network topologies. Application of our method to a broad set of empirical networks—spanning biological, technological, and ecological domains—uncovers consistent structural signatures associated with passing or blocking behavior. These findings shed light on how the network architecture and the particular selection of input and output nodes shape information flow in real-world systems, with broad implications for control, signal processing, and network design.
\end{abstract}

\maketitle

\textbf{Complex dynamical networks live within an extended environment, with which they are coupled through both input nodes that  receive stimulation signals and output nodes that release response signals. A fundamental question, which we address here with a focus on discrete time networks, is whether the network topology and the particular choice of input and output nodes determine an amplification or suppression in this input-to-output transformation. In this work, we focus on a single quantity: the $\mathcal{H}_2$-norm of an open network, which captures this input-output relationship. We derive a simple approximation formula for the $\mathcal{H}_2$-norm and introduce a normalized structural index that quantifies how the choice of input nodes influences the amplification or attenuation of signals. These concepts are evaluated on empirical networks from various domains.}

\section{Introduction}

Understanding complex networks in discrete time is essential for accurately modeling systems where interactions and updates occur at distinct intervals rather than continuously. 
Many real-world networks—including digital communication systems, social networks, and biological processes such as gene regulation or neural spike trains—operate through inherently discrete events or sampling procedures. Discrete-time models are not only more appropriate for systems governed by algorithmic or clock-driven mechanisms, but they also align naturally with data acquisition, which typically occurs in time series form. Moreover, the mathematical and computational frameworks for discrete systems enable tractable analysis, efficient simulation, and compatibility with digital hardware, making them indispensable for both theoretical studies and practical applications in control, prediction, and inference on networks.

This work centers on the study of \textit{open networks}---that is, networks that interact dynamically with their  environment by receiving inputs at certain nodes and transmitting outputs through others. Such interaction is characteristic of many real systems: neural circuits processing sensory inputs~\cite{horrocks2024flexible,xue2024theoretical}, electrical power grids that both receive energy from generators and deliver it to consumers~\cite{horrocks2024flexible,xue2024theoretical}, or ecological networks influenced by environmental fluctuations~\cite{meng2020tipping}. While extensive research has addressed control of networked systems (see, e.g.,~\cite{yan2015spectrum,gao2014target,klickstein2017energy,shirin2017optimal,iudice2019node,gambuzza2020controlling}, our focus is distinct: we investigate how networks respond to a wide range of environmental signals, beyond those strictly used for control purposes. {{For recent reviews of dynamical processes and signal propagation in complex networks, see \cite{artime2024robustness}  and \cite{ji2023signal}.}}

Prior studies have examined network responses to external drivers~\cite{tyloo2018robustness,young2010robustness,tyloo2023finite,zhang2019fluctuation} in continuous time, but these analyses often rely on simplifying assumptions---such as symmetric connectivity or, at most, asymmetric networks with normal adjacency matrices. However, many real networks are \textit{non-normal}, highly \textit{directional}, and exhibit layered or \textit{hierarchical organization}---features that significantly impact their dynamic behavior and are mathematically much harder to treat~\cite{johnson2017looplessness, asllani2018structure, asllani2018topological, MUOLO2019Patterns, muolo2020synchronization, duan2022network, nazerian2023Communications,lizier2023analytic,tyloo2025predicting}. {Inspired by these findings, information-theoretic results have been developed to study how flow and information propagate in directed networks \cite{o2021hierarchical,ramon2024entropy}.} 
In this paper, different from this previous work,  we present a \textit{general framework} for analyzing the response of directed networks to various input signals---ranging from sinusoidal and periodic to stochastic and broadband. The study here presented  parallels another work from some of the same authors which focuses on continuous-time open networks \cite{nazerian2025frequency}.

We define an open network through three components: (i) its internal structure, given by the node set $\mathcal N$ and a set $\mathcal E$ of directed, possibly weighted, edges; (ii) a subset of input nodes $\mathcal I \subseteq \mathcal N$ where signals from the environment are introduced; and (iii) a subset of output nodes $\mathcal O \subseteq \mathcal N$, where the resulting network response is observed. Our analysis examines how both the connectivity pattern and the placement of input/output nodes influence the \textit{attenuation or amplification} of incoming signals. We quantify this behavior via the network’s \textit{transfer function}, using the \textit{$\mathcal{H}_2$-norm} as a comprehensive measure of signal gain across various classes of inputs. Additionally, we establish a formal link between the network’s structural features, its transfer function, and the \textit{controllability Gramian}, which plays a pivotal role in our theoretical development. This framework enables us to determine whether a given network tends to \textit{transmit} or \textit{suppress} external stimuli, depending on the choice of input and output nodes.

Using this approach, we conduct a comparative analysis of empirical networks drawn from diverse domains, aiming to identify general structural features that govern how networks reshape and process external signals. We see that networks from different domains are typically characterized by distinct passing vs blocking behavior. 

\color{black}

\section{Analysis of Networks as Open Systems}

{Networks dynamics models in discrete time emerge in diverse contexts: they describe opinion updating in social systems through neighbor averaging, the evolution of probability distributions in Markov chains driven by a transition matrix, the diffusion of heat or mass across interconnected structures, and the progression of populations across age classes governed by the Leslie matrix \cite{leslie1945use,li2009outer,dubovskaya2023analysis,delvenne2015diffusion,nazerian2023reactability}. In each of these examples, the evolution of the system is determined by the current state and the structure of the network interactions encoded in the matrix \( A \). 
The dynamics of all these complex interconnected systems 
arise from the interplay between the internal node dynamics, the network interactions and the external signals. Near a stable equilibrium, these systems can often be approximated by a linear time-invariant model in discrete time (map), expressed as \( \mathbf{x}_{k+1} = A \mathbf{x}_k \), where \( \mathbf{x}_k \) represents the state of all nodes at time \(k\) and \( A \) encodes the interaction weights. }

{ Building on this insight, we now frame the dynamics as a standard linear time-invariant (LTI) discrete time systems, described by the equations}
\begin{subequations} \label{eq:discrete}
\begin{align}
    \bx_{k+1} & = A \bx_k + B \bu_k\label{eq:discrete1} \\
    \by_k & = C \bx_k,\label{eq:discrete2}
\end{align}
\end{subequations}
where $\bx_k \in \mathbb{R}^N$ is the network state vector, and $\bu_k \in \mathbb{R}^M$ is the vector of external input signals acting on the network, reflecting its open nature. The output vector $\by_k \in \mathbb{R}^P$ collects signals from selected nodes. 
The matrix $A \in \mathbb{R}^{N \times N}$ describes the network connectivity, i.e., $A_{ij}$ is the strength of the coupling from node $j$ to node $i$ ($A_{ij}=0$ if node $j$ does not directly affect node $i$.) The matrix $B \in \mathbb{R}^{N \times M}$ has $M$ non-repeated versors as columns, namely, if $B_{ij} = 1$, then $B_{kj} = 0$, $\forall k \neq i$. The matrix $C \in \mathbb{R}^{P \times N}$ has $P$ non-repeated versors as its rows, namely, 
 if $C_{ij} = 1$, then $C_{ik} = 0$,
$\forall k \neq i$. Our choice of the matrices $B$ and $C$ describes
which nodes are input and output nodes, respectively. If
$B_{ij} = 1$, it means that node $i$ is an input node, $i \in \mathcal I$ and receives
input signal $j$. Similarly, if $C_{ij} = 1$, it means that
node $j$ is an output node, $j \in \mathcal O$, and transmits output signal $i$.
We proceed under the assumption that all of the eigenvalues $\lambda_i$ of the matrix $A$ lie inside the unit circle.
In order to satisfy this assumption, we set 
\begin{equation} \label{eq:rescaling}
A \leftarrow \frac{\rho}{\max_{i} | \lambda_i|}A,
\end{equation}
where $0 < \rho <1$ is the desired spectral radius for the matrix $A$.

The $\mathcal{H}_2$ norm for the discrete-time linear time-invariant system in \eqref{eq:discrete} is
\begin{equation}
    \| G \|_2^2 \coloneq \sum_{k=0}^{+\infty} \| Y_k \|_F^2 = \Tr (C W_d C^\top).
\end{equation}
Here, $Y_k = C A^{k-1} B$ is the response to the input impulse at time $k$, and $\| \cdot \|_F$ is the Frobenius norm. 
The symmetric positive definite matrix $W_d$ is the discrete-time controllability Gramian and is defined as
\begin{equation}
     W_d = \sum_{j=0}^{\infty} A^j B B^\top ({A^\top})^j.
\end{equation}
The matrix $W_d$  satisfies the discrete-time Lyapunov equation:
\begin{equation} \label{eq:lyapunov}
    A W_d A^\top - W_d + B B^\top = 0,
\end{equation}
which is a linear system of equations in the entries of $W_d$ and can be solved easily.
We denote the output controllability Gramian as 
\begin{equation} \label{eq:wout}
    W_d^{out} = C W_d C^\top,
\end{equation}
where $W_d$ is obtained by solving Eq.\,\eqref{eq:lyapunov}.
\begin{remark}
    The average energy to move the system in state-space in $K$ steps is inversely related to the trace of the controllability Gramian:
    \begin{equation}
        \Tr(W_K) = \Tr\left( \sum_{j = 0}^{K-1} A^j B B^\top (A^\top)^j \right).
    \end{equation}
\end{remark}

An important property of the trace of the output controllability Gramian $\Tr(CW_dC^\top)$ is modularity  with respect to the columns of the matrix $B$ \cite{summers2015submodularity} and also with respect to the rows of the matrix $C$ \cite{nazerian2025frequency}.
This property implies that the trace of the output controllability Gramian is equal to the sum of the individual contributions of each one of the traces resulting from choosing each column of $B$ and each row of $C$, i.e., 
\begin{equation}
    \Tr(CW_dC^\top) = \sum_{i=1}^M \sum_{k=1}^P C_k W_d^{(i)} C_k^\top, 
\end{equation}
where $W_d^{(i)} = \sum_{j=0}^{\infty} A^j B_i B_i^\top ({A^\top})^j$, and $B_i$ and $C_k$ are column $i$ and row $k$ of the matrices $B$ and $C$, respectively.
In the case of interest here, where $B_i$ and $C_k$ have the versor structure, this implies that the overall trace is the sum of the contributions of each pair of  input-output nodes to the trace.
That is
\begin{equation}
    \Tr(W_d(A,\mathcal{I},\mathcal{O})) = \sum_{i \in \mathcal{I}} \sum_{k \in \mathcal{O}} \Tr(W_d(A,i,k))
\end{equation}
where $W_d(A,\mathcal{I},\mathcal{O})$ denotes the controllability Gramian defined by the matrix $A$ and the sets of input and output nodes $\mathcal{I}$ and $\mathcal{O}$, respectively. This is important as it allows us to study the general case of networks with multiple inputs and multiple outputs, in terms of the contributions of single-input single-output networks.

\color{black}

\subsection{Output Controllability Gramian Approximation}

{In this section, we derive an approximation for the trace of the output controllability Gramian, which corresponds to the squared \(\mathcal{H}_2\)-norm of the network, explicitly capturing the underlying network structure encoded by the triplet \((A, B, C)\). We focus on the case of a network with a single input and a single output, where \(B \in \mathbb{R}^N\) and \(C \in \mathbb{R}^{1 \times N}\), and we retain the assumption that both \(B\) and \(C\) have versor structure. Let \(d\) denote the distance—defined as the length of the unweighted shortest path—from the input node to the output node.}
{The output controllability Gramian is, 
\[
W_d^{out} := C W_d C^\top = \sum_{i=0}^\infty C A^i B B^\top (A^\top)^i C^\top.
\]  
Since we consider the case of a single-input and single-output, the terms in the summation are scalars, and the expression simplifies to  
\[
W_d^{out} = \sum_{i=0}^\infty \big( C A^i B \big)^2.
\]  
We note that the zero-order term \( (CB)^2 \) vanishes unless the input and output nodes coincide. In what follows, we assume that the input and output nodes are distinct.}
{Moreover, the terms corresponding to powers lower than \(d\) steps vanish, as no path exists from input to output in fewer than \(d\) steps. The summation can thus be rewritten as follows,
\begin{align}
W_d^{out} &= \sum_{i=0}^\infty \big( C A^{d+i} B \big)^2 \approx \sum_{i=0}^\infty \big( C A^d B \big)^2 \rho^{2i},
\end{align}
where we replaced the powers \( A^{2i} \) by \( \rho^{2i} \), where \( \rho \) (recall we set \( 0 < \rho < 1 \) ) is the spectral radius of \( A \).
For a sufficiently large input–output distance \(d\), \( A^{d+i} \) asymptotically scales as \( \rho^{d+i} \), since the largest eigenvalue dominates at large powers. The summation above, being a convergent geometric series in \( \rho^{2i} \), then can be rewritten as follows}
\begin{equation} \label{eq:Wapprox}
    \hat{W}_d^{out} \approx \dfrac{(C A^d B)^2}{(1-\rho^2)}.
\end{equation}
Note that the simple formula in Eq.\ \eqref{eq:Wapprox} depends on two important network structural quantities, the spectral radius $\rho$ and the  input node to the output node distance $d$.
Since the $\mathcal{H}_2$-norm squared is equal to the trace of the output controllability Gramian, and also based on the modularity property, then Eq.\,\eqref{eq:Wapprox} can be used multiple times for each pair of the input-output nodes of a multi-input-multi-output network to provide the following expression:
\begin{equation} \label{eq:Wapprox2}
    \Tr(\hat{W}_d^{out}(A,\mathcal{I},\mathcal{O})) = \sum_{i \in \mathcal{I}} \sum_{k \in \mathcal{O}} \hat{W}_d(A,i,k),
\end{equation}
where $\hat{W}_d(A,i,k)$ is the approximation of the output controllability Gramian using Eq.~\eqref{eq:Wapprox} when node $i$ is the input and node $k$ is the output.
We recall that the sets $\mathcal{I}$ and $\mathcal{O}$ denote the sets of input and output nodes, respectively.

\begin{figure}
    \centering
    \includegraphics[width=0.9\linewidth]{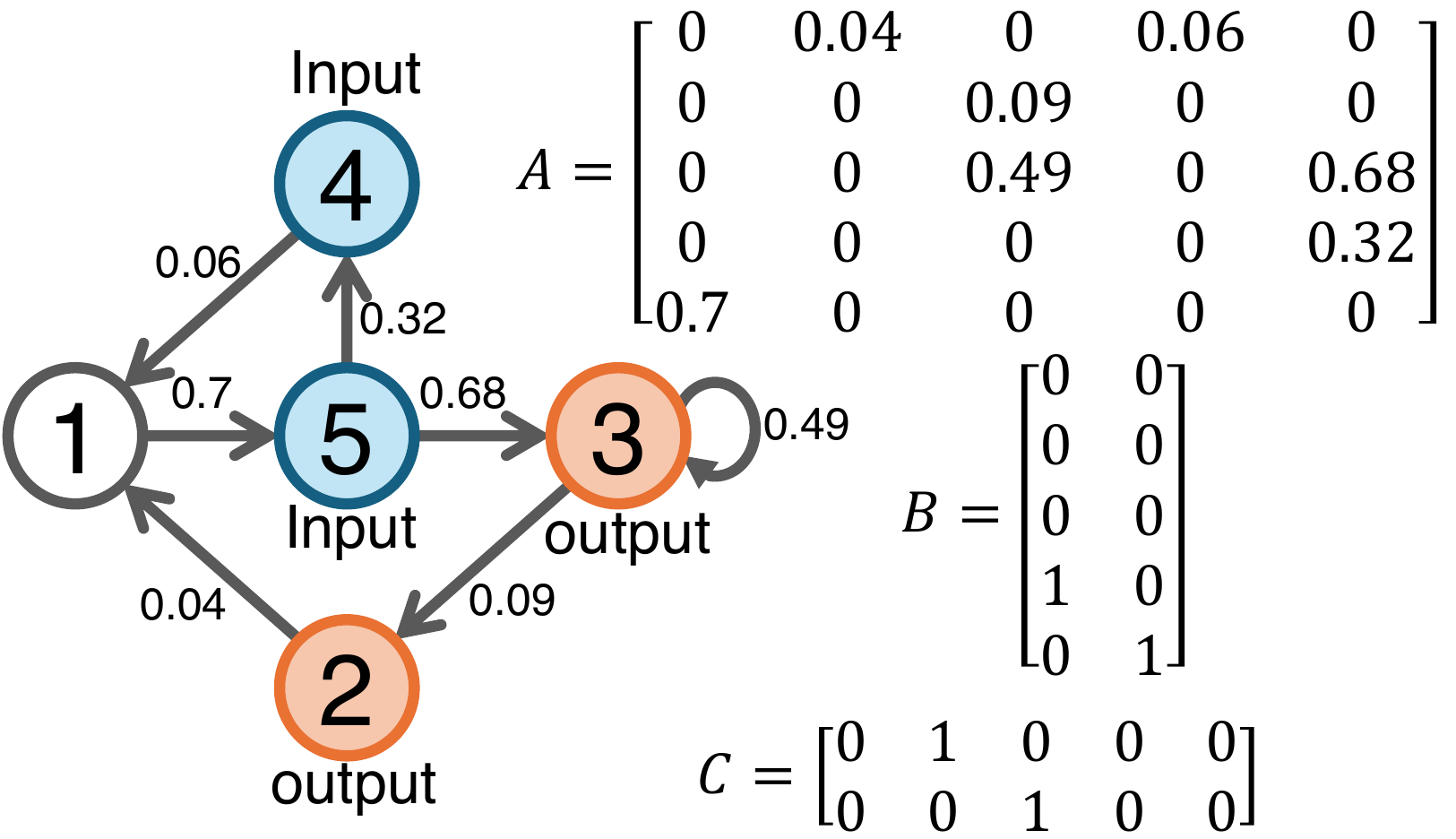}
    \caption{
    Randomly generated graph for Example 1, where the adjacency matrix $A$, the input matrix $B$, and the output matrix $C$ are provided {on the right side of}  the figure. {Input nodes are represented in blue color and output nodes in red color.}}
    \label{fig:network}
\end{figure}

\textbf{Example 1:} 
Here, we randomly generate a graph, shown in Fig.~\ref{fig:network}.
For this example, $\rho (A) = 0.5$.
Also, the set of input nodes is $\mathcal{I}=\{ 4,5\}$, and the set of output nodes is $\mathcal{O} = \{ 2,3\}$.
Table~\ref{tab:example} provides information on the contribution of each pair of input and output nodes, where we see that for shorter path lengths $d$, the contribution $\hat{W}_d^{out}(A,i,k)$ is larger.
The trace of the exact output controllability Gramian by solving Eq.\,\eqref{eq:lyapunov} is $\Tr(W_d^{out}) = \Tr(C W_d C^\top) = 0.6168$, where its approximation using Eq.\,\eqref{eq:Wapprox2} is $ \Tr(\hat{W}_d^{out}(A,\mathcal{I},\mathcal{O})) = 0.6267$.

We calculate the normalized error
\begin{equation} \label{eq:normerr}
    E = \dfrac{\left| \Tr(\hat{W}_d^{out})- \Tr(W_d^{out}) \right|}{\Tr(W_d^{out})},
\end{equation}
where $\hat{W}_d^{out}$ is given in Eq.~\eqref{eq:Wapprox} and $W_d^{out}$ is given in Eq.\,\eqref{eq:wout}.
Here, we obtain $E = 0.0161$. 
\color{black}

\begin{table}
    \centering
    \caption{
    Additional information for Example 1: $i\to j$ denotes the path from input node $i$ to the output node $j$, $d$ is the unweighted distance along this path, $\hat{W}_d^{out}(A,i,k)$ and $W_d^{out}(A,i,k)$ are the approximated and true contribution of this path toward the trace of the controllability Gramian in \eqref{eq:Wapprox2}, and \eqref{eq:wout}, respectively, and $E(A,i,k)$ is the normalized error in \eqref{eq:normerr}.}
    \begin{tabular}{p{1.25cm}p{0.5cm}p{1.95cm}p{1.95cm}p{1.1cm}}
        $i \to k$ & $d$ & $\hat{W}_d^{out}(A,i,k)$ & $W_d^{out}(A,i,k)$ & $E(A,i,k)$ \\
        \hline \hline
        $4 \to 2$ & 4 & $8.8671e-06$ & $8.7262e-06$ & $0.0159$ \\
        $4 \to 3$ & 3 & $0.0011$ & $0.0011$ & $0.0060$  \\
        $5 \to 2$ & 2 & $0.0050$ & $0.0049$ & $0.0200$ \\
        $5 \to 3$ & 1 & $0.6206$ & $0.6107$ & $0.0160$ \\
        \hline
    \end{tabular}
    \label{tab:example}
\end{table}

\color{black}

\subsection{Numerical analysis of network topologies}


To assess the accuracy of the proposed approximation, we conduct a numerical analysis on synthetic directed networks, comparing the true and approximated output controllability Gramians, following the same approach outlined above. We consider two classes of networks: directed Erdős–Rényi (ER) graphs and directed scale-free (SF) graphs. {Next we describe how we construct the matrix $A$ in the two cases of ER networks and SF networks.} 
In the ER case,  {$A$ is taken to be the adjacency matrix of a} directed and weighted random graphs of \(N\) nodes, where each possible directed edge from node \(i\) to node \(j\) is included independently with probability \(p\). Each existing {directed} edge is assigned a weight drawn uniformly at random from the interval \([0,1]\). The resulting adjacency matrix is then rescaled so that the spectral radius equals \(\rho = 0.5\), following Eq.\ \eqref{eq:rescaling}.

For the SF networks, {we take $A$ to be the adjacency matrix of a directed, unweighted scale-free graph with prescribed in- and out-degree distributions}, generated using the static model proposed by Goh \textit{et al.}~\cite{goh2001universal}. The power-law exponents of the in-degree and out-degree distributions are given by \(\gamma_{\mathrm{in}} = \frac{1+\beta_{\mathrm{in}}}{\beta_{\mathrm{in}}}\) and \(\gamma_{\mathrm{out}} = \frac{1+\beta_{\mathrm{out}}}{\beta_{\mathrm{out}}}\), where \(\beta_{\mathrm{in}}, \beta_{\mathrm{out}} \in (0,1]\)
are control parameters. Each node \(i\) is assigned weights \(p_i = i^{-\beta_{\mathrm{out}}}\) and \(q_i = i^{-\beta_{\mathrm{in}}}\) that determine the probabilities of selecting \(i\) as the source and target of an edge, respectively. At each step, two distinct nodes \(i\) and \(j\) are sampled with normalized probabilities \(p_i / \sum_k p_k\) and \(q_j / \sum_k q_k\), and a directed edge from \(i\) to \(j\) is added if it does not already exist. This process is repeated until \(\kappa N\) edges have been created  , where \(\kappa\) is the desired average degree of the directed graph. In our experiments, we set \(\beta = \beta_{\mathrm{in}} = \beta_{\mathrm{out}}\) and vary $\beta$ from \(0.5\) to \(1\) while fixing \(\kappa=0.02N\), producing directed and unweighted SF networks with \(N\) nodes and degree distribution exponent \(\gamma = \gamma_{\mathrm{in}} = \gamma_{\mathrm{out}}\). The choice \(\beta=0.5\) (\(\beta=1\)) corresponds to a more homogeneous (more heterogeneous) scale-free network with \(\gamma=3\) (\(\gamma=2\)). {Finally, similarly to what done for ER networks, the obtained adjacency matrix is rescaled using Eq.\ \eqref{eq:rescaling}  in order to set the spectral radius to \(\rho = 0.5\).}

To systematically evaluate the accuracy of the approximation \eqref{eq:Wapprox} in both Erdős–Rényi (ER) and scale-free (SF) networks, we proceed as follows. 
We select a single input node and a single output node, such that the path from the input node to the output node has (unweighted) length $d$ among all possible pairs of nodes. 
For each choice of $(d,p)$ in ER networks and each choice of $(d,\beta)$ in SF networks, we randomly select 100 pairs of \{input, output\} nodes.
For each pair of input-output nodes, we then create the matrices $B$ and $C$ with versor structures, accordingly.
We vary the size of the network $N$ and the connection probability $p$ (for ER networks) or the control parameter $\beta$ (for SF networks) and calculate the normalized error $E$.

\begin{figure*}
    \centering 
    \includegraphics[width=0.9\linewidth]{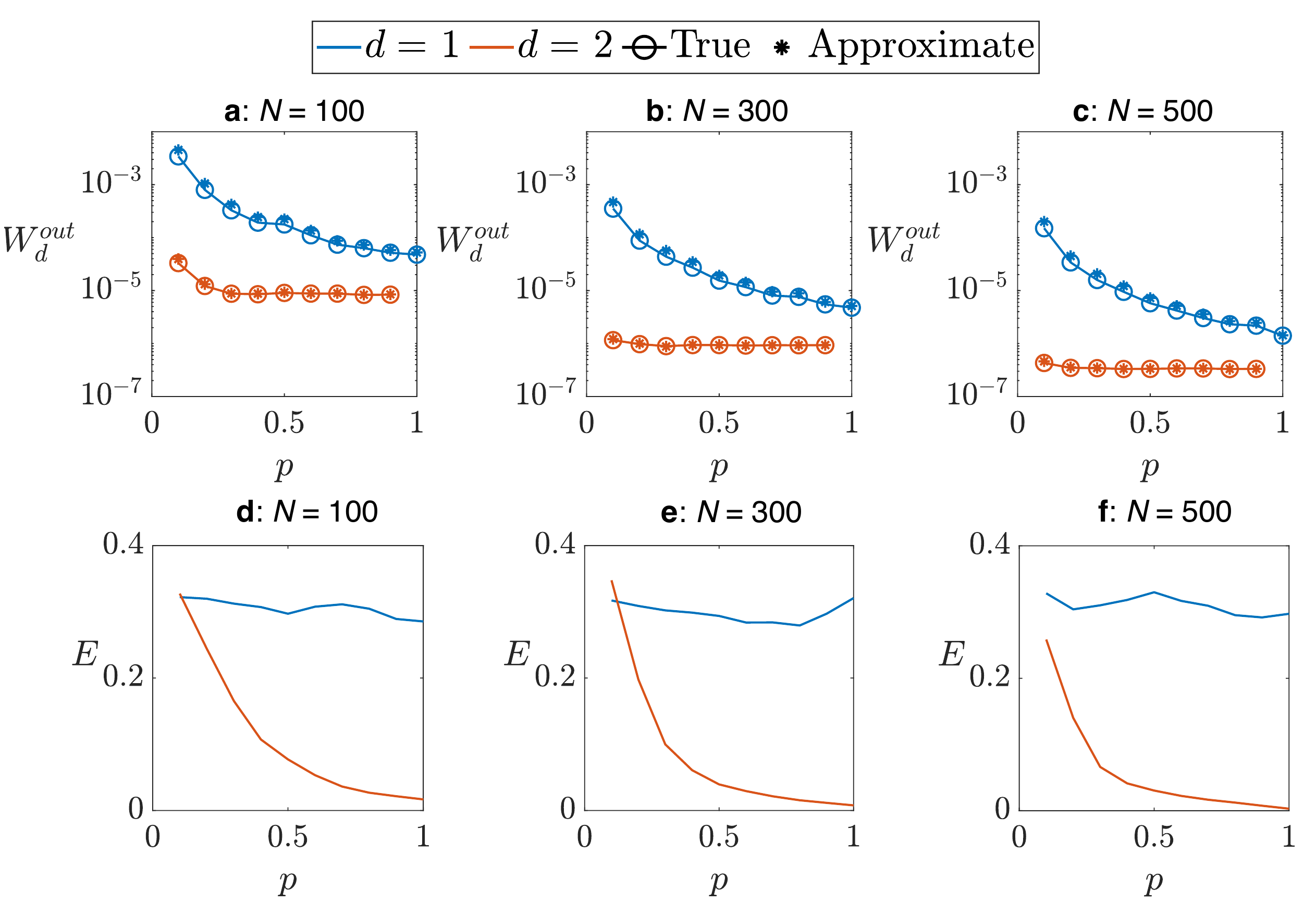}
    \caption{
    Panels a-c show a comparison between the true (Eq.\,\eqref{eq:wout}) and approximated (Eq.\,\eqref{eq:Wapprox}) output controllability Gramian. The blue color (red color) is for the case that the distance from the input to the output node is $d=1$ ($d=2$). Circles are true values and stars are approximated values. {We see that the output controllability Gramian $W_d^{out}$ decays for larger ER networks (larger $N$), for more dense networks (larger $p$), and for longer distances between the input and the output nodes (larger $d$). The figure only shows the two cases that the distance from the input node to the output node is either $d=1$ or $d=2$,  with the latter yielding a much better performance of the approximation than the former (see also the lower panels). Larger values of $d$, for which the approximation works even better are not shown.}
    Panels d-f show the normalized error $E$ (Eq.\,\eqref{eq:normerr}) as the connection probability $p$ of Erd{\H{o}}s-R{\'e}nyi graph is varied, for different numbers of nodes $N$. {We see that the error $E$ decays with the distance $d$.}
    The data is averaged over 100 realizations for each choice of $(d,p)$ in all panels.}
    \label{fig:ER}
\end{figure*}

Figure~\ref{fig:ER} a-c shows that the output controllability Gramian $W_d^{out}$ decays for larger ER networks (larger $N$) and for more dense networks (larger $p$), and for longer distances between the input and the output nodes (larger $d$). 
Figure~\ref{fig:ER} d-f shows that the approximation of the output controllability Gramian \eqref{eq:Wapprox} is more accurate for larger ER networks (larger $N$) and for more dense networks (larger $p$), and for longer distances between the input and the output nodes (larger $d$). We note that the approximation matches well the true output controllability Gramian in all cases, with small discrepancy.

\begin{figure*}
    \centering 
    \includegraphics[width=0.9\linewidth]{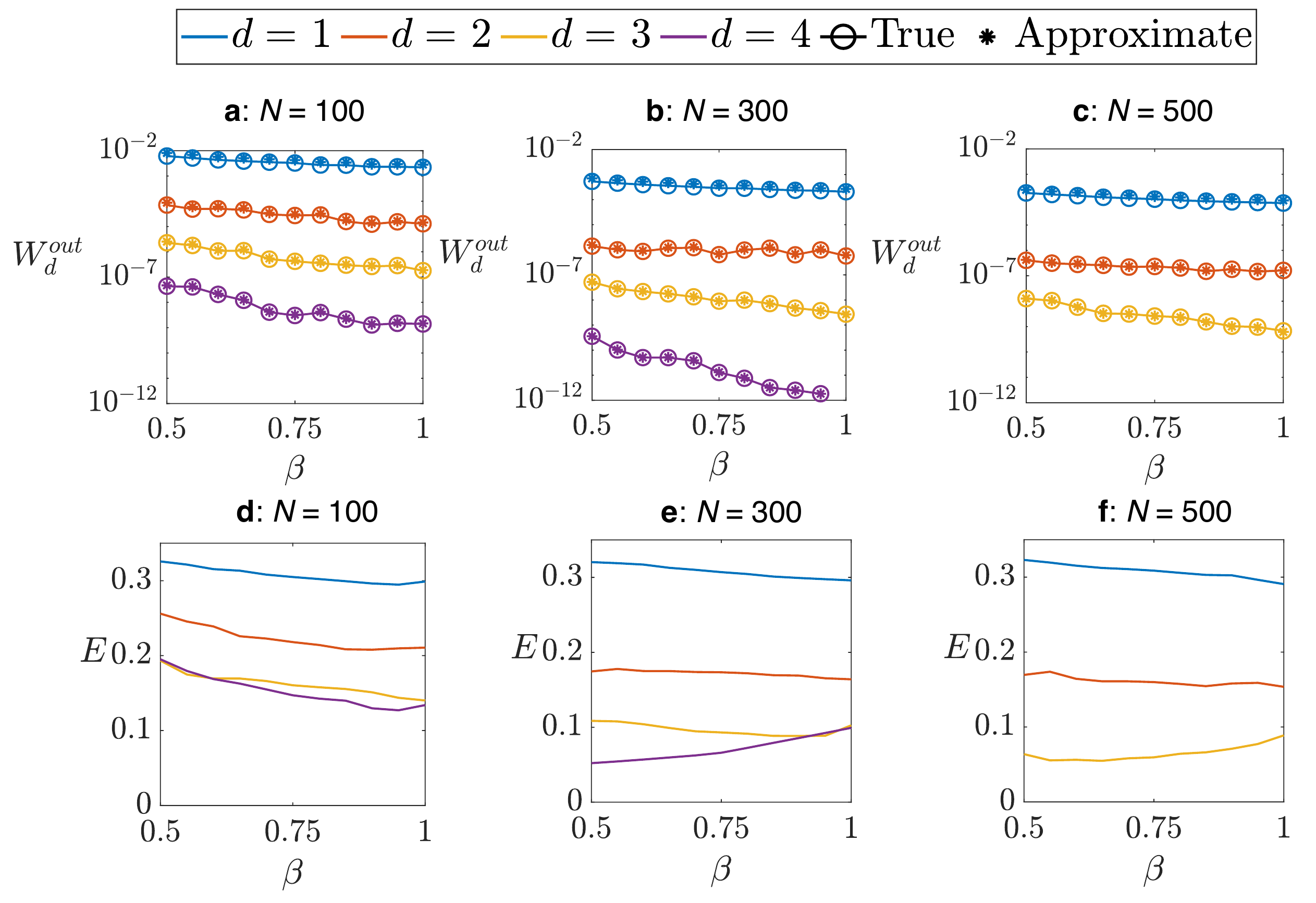}
    \caption{Panels a-c show a comparison between the true (Eq.\,\eqref{eq:wout}) and approximated (Eq.\,\eqref{eq:Wapprox}) output controllability Gramians for different values of the path length $d$ from the input to the output node, {as the control parameter $\beta$ of the scale-free network is varied, for different numbers of the network nodes $N$. The degree distribution of the scale free networks is more homogeneous for larger values of $\beta$. For example, the choice \(\beta=0.5\) (\(\beta=1\)) corresponds to a more homogeneous (more heterogeneous) scale-free network with \(\gamma=3\) (\(\gamma=2\)). } Circles are true values and stars are approximated values.
    Panels d-f show the normalized error $E$ (Eq.\,\eqref{eq:normerr})  as the control parameter $\beta$ of the scale-free network is varied, for different numbers of nodes $N$. {In all panels we see that the error decreases with the distance $d$.} 
    The data is averaged over 100 realizations for each choice of $(d,\beta)$ in all panels.
}
    \label{fig:SF}
\end{figure*}

Figure~\ref{fig:SF} a-c shows that the output controllability Gramian decays for larger SF networks (larger $N$) and for more homogeneous networks (larger $\beta$), and for longer distances between the input and the output nodes (larger $d$). 
Figure~\ref{fig:SF} d-f shows that the approximation \eqref{eq:Wapprox} of the output controllability Gramian is more accurate for larger SF networks (larger $N$) and for more homogeneous networks (larger $\beta$), and for longer distances between the input and the output nodes (larger $d$). {In Fig.~\ref{fig:SF} e, the increase in the error is due to the finite machine precision.}
The approximation matches well the true output controllability Gramian in all cases, with small discrepancy.

\color{black}

{\section{Structural analysis and empirical validation}}

\subsection{The Network Index $\alpha$}

\begin{figure}
    \centering
    \includegraphics[width=0.65\linewidth]{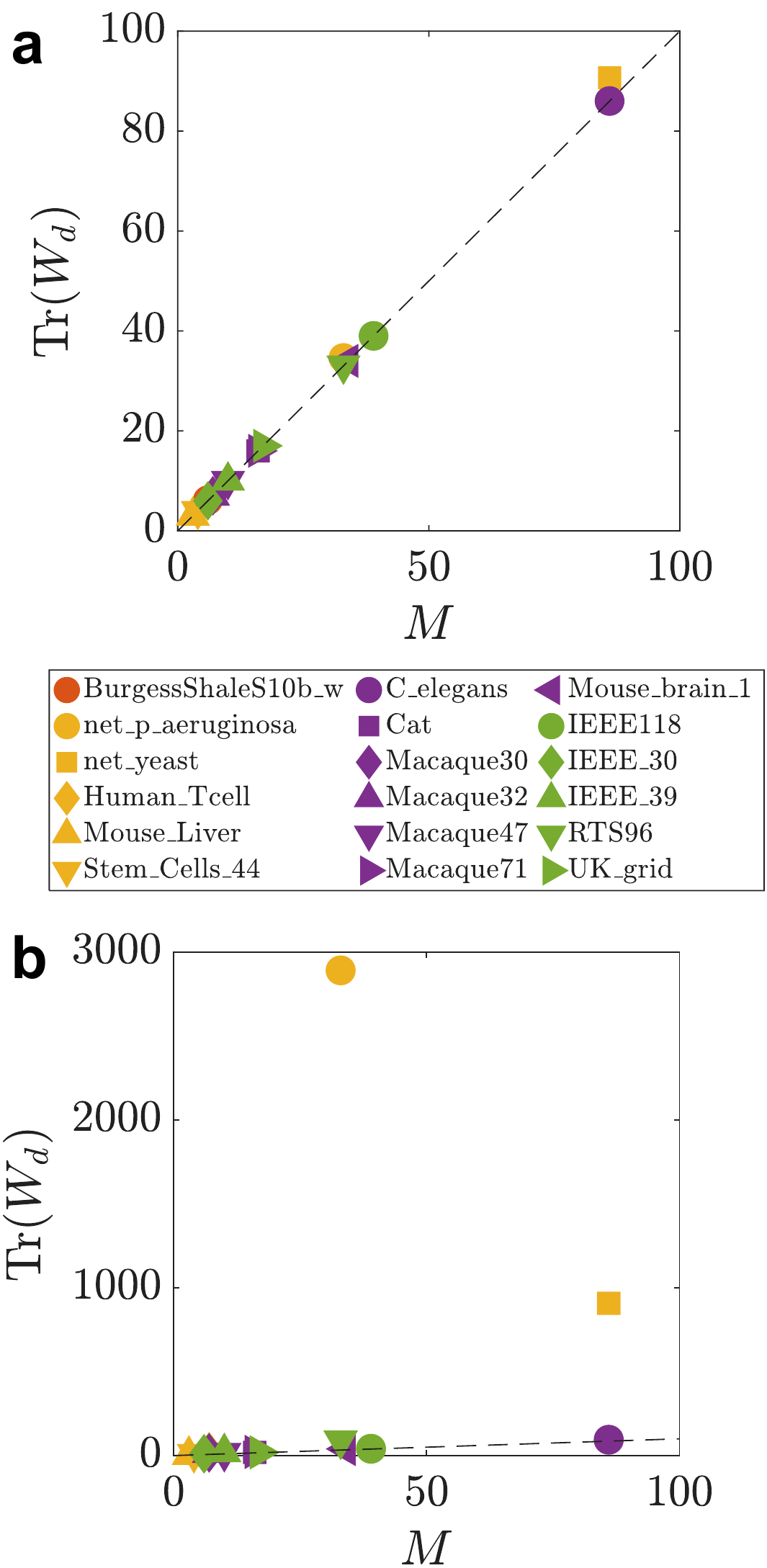} 
    \caption{
    The trace of the infinite-horizon discrete-time controllability Gramian $\Tr(W_d)$ for different real networks vs their number of input nodes $M$.
    In panels a and b, the adjacency matrices are scaled such that all networks have spectral radii of 0.1 and 0.98, respectively. The dashed black line represents $\Tr(W_d) = M$.  {We see that when the spectral radii of the adjacency matrices are small (panel a), $\Tr(W_d) \approx M$, and when the spectral radii are large (panel b), $\Tr(W_d) \geq M$.}}
    \label{fig:trace_pinned}
\end{figure}

{
Here, we aim to provide a normalized structural network index that quantifies whether the selection of $M$ input nodes for a network results in the lowest/highest possible trace of the output Controllability Gramian ($\mathcal{H}_2$-norm).
}
Hereafter, without loss of generality, we assume nodes $1,2,...,M$ are the input nodes. Note that if this is not the case, the nodes and the input signals may be renamed to satisfy this assumption. As a result, we have the matrix
\begin{equation*}
    B = \begin{bmatrix}
        I_M  \\
        0_{N-M,M}
    \end{bmatrix}.
\end{equation*}
We denote the set of output nodes as $\mathcal{O} = \{ o_1, o_2, \hdots, o_P \}$.
The $P \times N$ matrix $C$ has all zero entries except for $C_{j,o_j} = 1$, $j = 1, \hdots, P$.
The $K$-step discrete-time output controllability Gramian is
\begin{equation*}
    C W_K C^\top = C \sum_{k=0}^{K-1} A^k B B^\top \left(A^\top \right)^k C^\top.
\end{equation*}
The trace of the output Gramian $C W_K C^\top$ can be calculated as
\begin{align*} \label{eq:sum_trace}
\begin{split}
    \Tr(C W_K C^\top) & = \Tr \left( C\sum_{k=0}^{K-1} A^k B B^\top \left(A^\top \right)^k C^\top\right) \\
    & = \Tr( CB B^\top C^\top) \\
    & \quad \ +\sum_{k=1}^{K-1} \Tr \left( C A^k B B^\top \left(A^\top \right)^k C^\top \right).
\end{split}
\end{align*}
It follows that term $k$ in the summation is
\begin{align*}
\begin{split}
    A^k B & = \begin{bmatrix}
    [A^k]_{11} & \hdots & [A^k]_{1m}\\ 
    [A^k]_{21} & \hdots & [A^k]_{2m} \\ 
    \vdots & \ddots & \vdots\\ 
    [A^k]_{N1} & \hdots & [A^k]_{Nm}
    \end{bmatrix}, \\
    B^\top \left(A^\top \right)^k & = \begin{bmatrix}
    [A^k]_{11} & [A^k]_{21} & \hdots & [A^k]_{N1}\\ 
    \vdots & \vdots & \ddots & \vdots  \\ 
    [A^k]_{1m} & [A^k]_{2m} & \hdots & [A^k]_{Nm} 
    \end{bmatrix}
\end{split}
\end{align*}
If the matrix $A$ is an unweighted adjacency matrix with binary entries, then $[A^k]_{ij}$ is the number of paths from node $i$ to $j$ in $k$ steps.
If the matrix $A$ is a weighted adjacency matrix, then $[A^k]_{ij}$ is the commutative sum of the weights of all $k$ step paths from node $i$ to $j$.
Thus, the trace of term $k$ in the summation is
\begin{equation*}
    \Tr \left(  A^k B B^\top \left(A^\top \right)^k \right) = \sum_{j=1}^M \sum_{o \in \mathcal{O}} [A^k \circ A^k]_{oj},
\end{equation*}
where $\circ$ denotes the Hadamard product (entrywise product).
Finally, the trace of the $K$-step discrete-time controllability Gramian is
\begin{equation}
     \Tr(W_K) = \Tr( CB B^\top C^\top) + \sum_{k=1}^{K-1} \sum_{j=1}^M \sum_{o \in \mathcal{O}} [A^k \circ A^k]_{oj}.
\end{equation}
If the spectral radius of the matrix $A$ is small, the trace will be dominated by $\Tr( CB B^\top C^\top)$, i.e., $\Tr(W_K) \approx \Tr( CB B^\top C^\top)$.
For larger spectral radii, the terms inside the summation will behave as positive perturbations to the base value $\Tr( CB B^\top C^\top)$.
An example of this is shown in Fig.~\ref{fig:trace_pinned}, where $\Tr(W_d)$ is evaluated for selected real networks. 
In this example, $C = I$, so $\Tr( CB B^\top C^\top) = \Tr( B B^\top) = M$.
We see that when the spectral radii of the adjacency matrices are small (panel a), $\Tr(W_d) \approx M$, and when the spectral radii are large (panel b), $\Tr(W_d) \geq M$.

Here, we separate the first term of the summation from the rest:
\begin{equation}
     \Tr(W_K) = M + \sum_{j=1}^M\sum_{i=1}^N A_{ij}^2 + \sum_{k=2}^{K-1} \sum_{j=1}^M \sum_{i=1}^N [A^k \circ A^k]_{ij}.
\end{equation}
Next we focus on the leading sum $\sum_{j=1}^M\sum_{i=1}^N A_{ij}^2$, based on which we define the network index $\alpha$ to provide a more computationally efficient metric to compare the trace of the controllability Gramian when different nodes of a network are selected as input nodes.


Then we introduce $d_j \coloneqq \sum_{i=1}^N A_{ij}^2$, of node $j$, $j=1,\hdots,N$, and, without loss of generality, assume the nodes are ordered such that nodes with lower index $j$ have lower $d_j$, i.e., $d_1 \leq d_2 \leq \cdots \leq d_N$.
We define the network index
\begin{equation} \label{eq:netindex}
    \alpha = \dfrac{\sum_{i=1}^M d_{i^*} - d_i}{\sum_{i=1}^M d_{N-M+i} - d_i},
\end{equation}
$0\leq \alpha \leq 1$, where $M$ is the total number of input nodes, and $i^*$ is the index of input node $i$ with the lowest $d_{i^*}$.
If $i^* = i, \  i =1,\hdots, M,$ then $\alpha =0$, which corresponds to the case in which the M input nodes are selected to be the ones with the lowest $d_i$ values.
If $i^* = N-M+i, \ i = 1, \hdots, M,$ then $\alpha = 1$, which corresponds to the case in which the M input nodes are selected to be the ones with the largest $d_i$ values.

For example, assume a network with $N=5$ nodes and $d_1 = 1, d_2 = 2, d_3 = 2, d_4 = 3, d_5 = 3$. If nodes 2 and 4 are input nodes, then $M=2$, $1^*=2$, and $2^* =4$.
The network index becomes
\begin{equation*}
    \alpha = \dfrac{(d_2 - d_1) + (d_4 - d_2)}{(d_4 - d_1) + (d_5 - d_2)} = \dfrac{(2-1) + (3-2)}{(3-1) + (3-2)} = \dfrac{2}{3}.
\end{equation*}
Evaluation of $\alpha$ is much more computationally efficient than evaluating the trace of the controllability Gramian, especially in large-scale networks. 
In what follows, we calculate $\alpha$ for several empirical networks and show that $\alpha$ and the trace of the Controllability Gramian are linearly correlated as long as the spectral radius of the adjacency matrix $A$ is small (less than 0.2 based on our numerical simulations).

\subsection{Optimal Selection of the Input Nodes}

An important question that we address in this paper is how to select $k$ input nodes so that the trace of the controllability Gramian is maximized/minimized.
Through the modularity property of the trace, the trace is recorded when each one of the nodes is set as the only input node. Then, $k$ nodes that provide the maximum/minimum trace are selected as the optimal set choice. 
If the initial network is such that the spectral radius of the adjacency matrix $A$ is small, then instead of evaluating the trace for each input node, it is sufficient to calculate $d_j$ for each node and select $k$ nodes that result in maximum/minimum values of $d_j$. 


\subsection{Optimization Applied to Real Networks}

We analyze empirical networks across various domains. {For all cases, $A$ is taken as the adjacency matrix of an empirical network reconstructed from publicly available datasets. The obtained adjacency matrices are then rescaled using Eq.\ \eqref{eq:rescaling} in order to set the spectral radius to \(\rho = 0.2\).} In each case, we identify input nodes as those through which energy, matter, or information flows into the network from the environment.
For example, in food webs plants and phytoplankton serve as entry points by transforming sunlight into biomass~\cite{krause2003compartments, dunne2008compilation, johnson2017looplessness};
in electrical power grids, generators supply energy to the system~\cite{athay1979practical, simonsen2008transient, delabays2023locating};
in brain networks (connectomes), sensory or afferent regions act as receivers of environmental stimuli~\cite{felleman1991distributed, young1993organization, scannell1999connectional, honey2007network, varshney2011structural, oh2014mesoscale};
in gene regulatory systems, upstream transcription factors are activated by external signals and control downstream gene expression~\cite{xu2013escape, barah2016transcriptional, johnson2017looplessness, fang2021grndb, weinstock2024gene};
and in signaling pathways, membrane-bound receptors or intracellular sensors perceive signals from the environment and initiate downstream responses~\cite{Laboratories_2022}.  Detailed information on each dataset we consider is provided in the table in the Appendix. For the sake of simplicity and due to the absence of more detailed information, in what follows we set $C=I$.

It is reasonable to assume that many biological networks have evolved to strike an optimal balance between transmitting and suppressing external inputs. For instance, some systems may need to enhance specific signals while filtering out unwanted noise. In the context of technological applications, an important challenge lies in adapting or redesigning existing networks to improve their ability to transmit desired signals while blocking irrelevant or harmful ones.
Next, we study the level of blocking versus passing behavior of several empirical networks.
For each dataset, we fix the number of input nodes $M$ to its empirical value. 
We first randomly choose 10,000 sets of $M$ nodes (with a uniform probability of selection) and evaluate $\Tr(W_d)$, the trace of the infinite-horizon discrete-time controllability Gramian.
We compare $\Tr(W_d)$ with the network index $\alpha$ based on the input nodes provided in the dataset.
We also perform an optimization to minimize $\Tr(W_d)$
by choosing $M$ optimal input nodes.
Specifically, given the topology (the $N$-dimensional matrix $A$) and the number of input nodes $M$, we solve the following mixed-integer linear program via Gurobi \cite{gurobi} and YALMIP \cite{yalmip} in MATLAB \cite{MATLAB}:
\begin{subequations}
\begin{align}
    \min_{b_1, b_2, \hdots, b_N} \quad & \Tr (W_d) \\
    \text{subject to} \quad &  W_d - A W_d A^\top - \text{diag} (b_1, \hdots, b_N) = 0 , \\
    & \sum_{i=1}^N b_i  = M, \\
    & b_i \in \{0, 1 \}, \quad i = 1, \hdots, N.
\end{align}
\end{subequations}

\begin{figure}
    \centering
    \includegraphics[width=\linewidth]{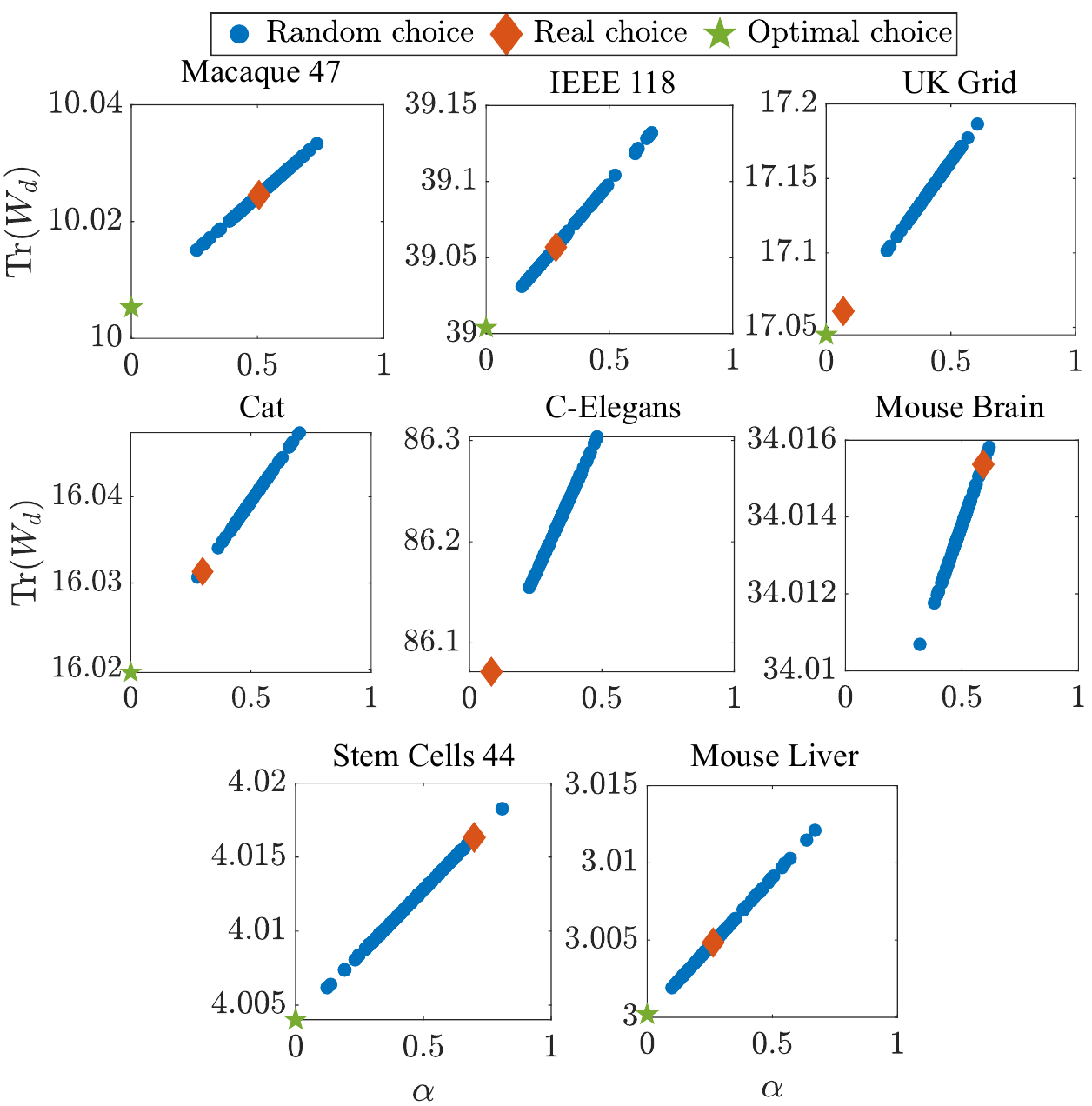}
    \caption{
    The trace of the infinite-horizon discrete-time controllability Gramian $\Tr(W_d)$ vs the network index $\alpha$ for selected real network datasets (IEEE118, UK Grid, Cat, C-Elegans, Mouse Liver, Mouse Brain, Stem Cells 44, and Macaque 47). For all datasets, the spectral radius of the adjacency matrix is scaled to be $\rho=0.2$. Each dataset has its own unique number of input nodes $M$. We randomly choose 100 sets of $M$ input nodes and evaluate $\Tr(W_d)$ and $\alpha$ for each set (blue circles). The pair value of $\Tr(W_d)$ and $\alpha$ from the input nodes within each real dataset is plotted as a red diamond. In each panel, a green star is used to label the optimal selection of  $M$ nodes, as described in the text. Due to the large size of the datasets C-Elegans and Mouse Brain, we were not able to perform the optimization.  {In all panels, we see an approximately linear relationship between the network index $\alpha$ and the trace of the (output) controllability Gramian, which is equal to the $\mathcal{H}_2$-norm squared. }}
    \label{fig:realrandomtrace}
\end{figure}

We then compare the values of $\Tr(W_d)$ for these 3 cases (with the same number $M$):  real data choice, random choice, and the optimal choice. {We call a network passing (blocking) when the empirical values of $\Tr(W_d)$ and $\alpha$ are higher (lower) than most random choices.}
In Fig.~\ref{fig:realrandomtrace}, we present the results of this analysis in the $\mathrm{Tr}(W_d)$--$\alpha$ plane for a variety of empirical networks (see Table II for additional information on the datasets.) Across all panels, we observe an approximately linear relationship between the network index $\alpha$ and the trace of the output controllability Gramian, $\mathrm{Tr}(W_d)$, which is equal to the square of the $\mathcal{H}_2$-norm. This relationship indicates that $\alpha$ serves as a robust proxy for comparing the signal amplification capacity of different networks, without requiring the explicit computation of the $\mathcal{H}_2$-norm or the Gramian itself.

{Overall, our analysis reveals a diverse range of behaviors across the empirical networks examined. In particular, the IEEE 118 network, the UK power grid network, the cat brain network, and the \textit{C. elegans} connectome consistently exhibit blocking behavior, in which input signals are attenuated rather than propagated. In contrast, the mouse brain network and the Stem Cells 44 network demonstrate passing behavior, with inputs being sustained or amplified. The remaining networks display intermediate or mixed responses, indicating that signal propagation capacity is not a uniform property but rather depends sensitively on the underlying organization and interaction structure of each network. Taken together, these findings suggest that technological networks often tend to exhibit more blocking behavior, while many biological networks are more often passing, although with notable exceptions. In particular, the contrasting behavior observed among neural networks from different species highlights that even within the same functional domain, signal propagation capacity can vary substantially, reflecting deeper differences in network architecture and functional organization. }
\color{black}

\subsection{Largest Eigenvalue of the Gramian}

From Eq. \eqref{eq:discrete1} we can write, 
\begin{equation} \label{eq:soldis2}
    \bx_{k} = \sum_{i = 0}^{k-1} A^{k-1-i} B \,\bu_i, \quad k \rightarrow \infty,
\end{equation}
We now focus on the effect of the $q<k$ most recent inputs $\bu_{k-1}, \bu_{k-2}, \hdots, \bu_{k-q}$ on the state $\bx_k$.
For this, we rewrite Eq.\,\eqref{eq:soldis2} as
\begin{align} \label{eq:soldis3}
\begin{split}
    \bx_{k} = &  \sum_{i = 0}^{k-(q-1)} \Big( A^{k-(q-1)-i} B \bu_i + B \bu_{k-1} \\
    & \quad + A B \bu_{k-2} + \hdots + A^{q-1} B \bu_{k-q} \Big), \quad k \rightarrow \infty,
\end{split}
\end{align} 
where we have isolated the last $q$ terms $A^{j-1} B \bu_{k-j}, \ j = 1, \hdots, q$, on the right-hand side of Eq.\,\eqref{eq:soldis3} to study the effect of the series of inputs $\bu_{k-1}, \bu_{k-2}, \hdots, \bu_{k-q}$ on the state $\bx_{k}$.
Assuming $q$ is large enough, then these terms have a greater effect on the norm of $\bx_k$ than the rest of the terms inside the summation.
Then we can rewrite Eq.\,\eqref{eq:soldis3} using matrix-vector notation as
\begin{equation*} \label{eq:soldis4}
    \bx_{k} = \sum_{i = 0}^{k-(q-1)} A^{k-(q-1)-i} B \bu_i + \mathcal{C}_q
    \bU_{k-1}^{(q)}
    \quad k \rightarrow \infty,
\end{equation*} 
where $\mathcal{C}_q = \left[B \ A B \ \cdots \ A^{q-1} B \right]$ and $\bU_{k-1}^{(q)} = \left[ \bu_{k-1}^\top \ \bu_{k-2}^\top \ \cdots \ \bu_{k-q}^\top \right]^\top \in \mathbb{R}^{pm}$.
If $q  = n$, then $\mathcal{C}_q$ is equal to the Kalman controllability matrix.
Thus we define the gain
\begin{equation}
    G = \sup_{\bU_{k-1}^{(q)}} \dfrac{\|\mathcal{C}_q \bU_{k-1}^{(q)} \|}{\| \bU_{k-1}^{(q)} \|} = \sigma_{\max} ( \mathcal{C}_q) = \sqrt{\lambda_{\max} (W_q) } 
\end{equation}
where the finite-horizon Gramian 
\begin{equation}
    W_q = \sum_{j=0}^{q-1} A^j B B^\top ({A^\top})^j.
\end{equation}

In the case that the measurable outputs of the system at time $k$ are $\by_k = C \bx_k \in \mathbb{R}^q$, the gain becomes
\begin{align}
\begin{split}
    G_{out} & = \sup_{\bU_{k-1}^{(q)}} \dfrac{\|C\mathcal{C}_q \bU_{k-1}^{(q)} \|}{\| \bU_{k-1}^{(q)} \|} \\
    & = \sigma_{\max} ( C\mathcal{C}_q) \\
    & = \sqrt{\lambda_{\max} (C W_q C^\top) } =: \sqrt{\lambda_{\max} ( W_q^{out} ) }
\end{split}
\end{align}
where the matrix $W_q^{out} \in \mathbb{R}^{q \times q}$ is the output controllability Gramian and is given by
\begin{equation}
    W_q^{out} = \sum_{j=0}^{q-1} C A^j B B^\top ({A^\top})^j C^\top.
\end{equation}

The larger the gain $G$ ( $G_{out}$), the larger the  amplification of the state $\bx_k$ (the output $\by_k$) from a series of critical inputs ${\bU^{(q)}_{k-1}}^*$, where ${\bU^{(q)}_{k-1}}^*$ is the right singular vector of $\mathcal{C}_q$ corresponding the largest singular value of $\mathcal{C}_q$.
In the limit in which $q \rightarrow \infty$,  $W_q$ becomes the controllability Gramian.

We repeat the process described in Sec.\ C and evaluate the largest eigenvalue of the infinite-horizon discrete-time controllability Gramian $\lambda_{\max}(W_d)$ for the following three selections of input nodes: 1- empirical data choice, 2- random choice, and 3- optimal choice corresponding to the case in which $\lambda_{\max}(W_d)$ is minimized. In all three cases, $M$ is kept fixed and equal to its empirical value.
Specifically, given the topology (the $N$-dimensional matrix $A$) and the number of input nodes $M$, we solve the following mixed-integer semi-definite program via Gurobi \cite{gurobi} and YALMIP \cite{yalmip} in MATLAB \cite{MATLAB}:
\begin{subequations}
\begin{align}
    \min_{b_1, b_2, \hdots, b_N, t} \quad & t \\
    \text{subject to} \quad & t I - W_d \succeq 0 \\
    & W_d - A W_d A^\top - \text{diag} (b_1,b_2, \hdots, b_N) = 0 , \\
    & \sum_{i=1}^N b_i  = M, \\
    & b_i \in \{0, 1 \}, \quad i = 1, \hdots, N.
\end{align}
\end{subequations}
\color{black}

\begin{figure}
    \centering
    \includegraphics[width=\linewidth]{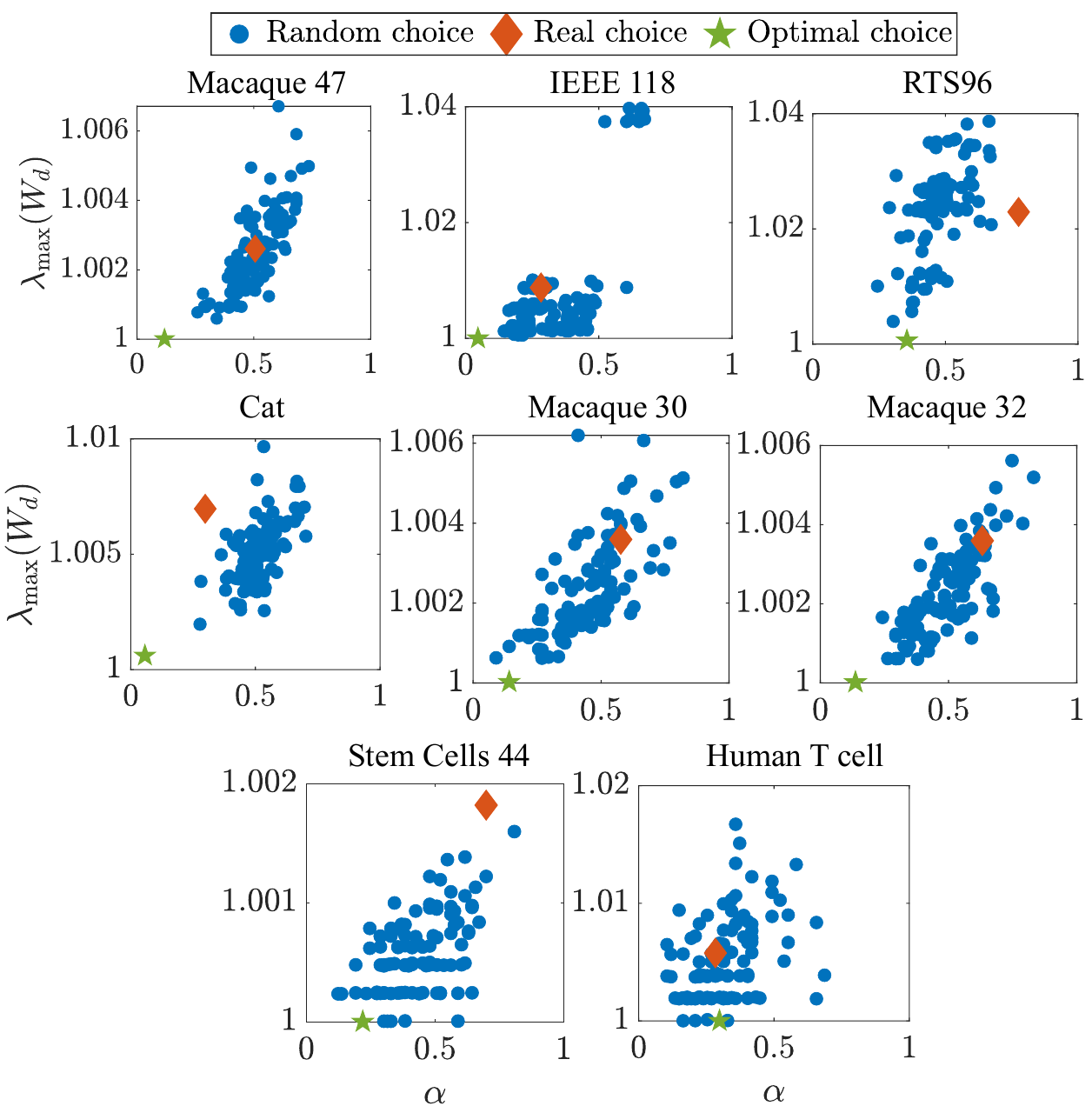}
    \caption{
    The largest eigenvalue of the infinite-horizon discrete-time controllability Gramian $\lambda_{\max}(W_d)$ vs the network index $\alpha$ for selected real networks (IEEE118, Macaque 47, Cat, Stem Cells 44, Macaque 32, RTS96, Human T cell, and Macaque 30). For all datasets, the spectral radius of the adjacency matrix is scaled to be $\rho=0.2$. Each dataset has its own unique number of input nodes $M$. We randomly choose 100 sets of $M$ input nodes and evaluate $\lambda_{\max}(W_d)$ and $\alpha$ for each set (blue circles). The pair value of $\lambda_{\max}(W_d)$ and $\alpha$ from the input nodes within each real dataset is plotted as a red diamond. In each panel, a green star is used to label the optimal selection of  $M$ nodes, as described in the text. }
    \label{fig:realrandomlambda}
\end{figure}

Figure \ref{fig:realrandomlambda} illustrates our results, by plotting he largest eigenvalue of the infinite-horizon
discrete-time controllability Gramian $\lambda_{\max}(W_d)$ vs the
network index $\alpha$ for selected real datasets.
{We often observe a positive correlation between $\lambda_{max}(W_d)$ and $\alpha$. Notably, in none of the networks does the empirical parameter choice lie close to the optimal one. Only in the case of the Stem Cells 44 network does the empirical configuration clearly exhibit passing behavior (which is in accordance with what seen in Fig.~\ref{fig:realrandomtrace}).
}

\section{Conclusions}

In this work, we introduced a comprehensive framework for analyzing the input-output response of discrete-time complex networks treated as open systems. Our approach leverages the network transfer function and the discrete-time controllability Gramian, with the $\mathcal{H}_{2}$-norm serving as a unified measure of signal amplification. Different from previous work, we do not restrict the analysis to symmetric or normal connectivity structures, which allows us to investigate the effects of the intrinsic asymmetries and non-normality present in real-world networks. 

We provide a simple formula (Eq. \eqref{eq:Wapprox}) to approximate the trace of the output controllability Gramian,  which critically depends on two important network structural quantities, the spectral radius $\rho$ and the  input node to output node distance $d$. {In particular, these two parameters appear explicitly in our formula Eq. \eqref{eq:Wapprox}, which has been found to well approximate the output controllability Gramian $W^{\text{out}}$ numerically. At the same time, our Eq. \eqref{eq:Wapprox} also depends explicitly on the adjacency matrix $A$, and therefore incorporates structural features such as the presence and placement of hubs. As a result, our analysis reflects all properties that can be inferred from the adjacency matrix. }

Another  important outcome of this paper is the  network index $\alpha$, which provides a computationally efficient proxy for estimating how the choice of
input nodes influences the amplification or attenuation of signals. We showed that the index $\alpha$ correlates well with the trace and the largest eigenvalue of the Gramian in networks with small spectral radius. This allows for scalable comparison across large datasets.

We applied our framework to a wide array of empirical networks—including biological, technological, and ecological systems—demonstrating that networks from different domains exhibit different passing or blocking tendencies. Our optimization analyses showed how modifying the choice of input nodes can significantly alter a network’s dynamical response. In particular, we found that real networks often deviate from random configurations and reflect structure-function relationships that may arise from evolutionary or design constraints. {Our work complements~\cite{o2021hierarchical, ramon2024entropy} which showed that increasing directionality/non-normality corresponds to a lower entropy rate (less dispersion and stronger channeling) and that broken detailed balance is captured by higher entropy production (irreversible one-way transport),  but also targets directional input–output transfer, providing a concise framework to distinguish transmission from blocking.} These results offer a general method for assessing and tuning signal flow in complex systems, with broad implications for network control, biological signal processing, and the design of robust engineered systems.

This work parallels another paper \cite{nazerian2025frequency} 
that focuses on continuous-time open networks. The main results of this discrete time paper, which are not found in Ref. \cite{nazerian2025frequency} are the network index $\alpha$ in Eq. \eqref{eq:netindex} and the approximation for the output controllability Gramian in Eq. \eqref{eq:Wapprox}. 

\appendix

{\section{Description of the empirical networks}}

The summary of the information on real networks is provided in Table \ref{tab:data}.

\begin{table*}
    \centering
    \caption{Real networks information.}
\resizebox{0.95\linewidth}{!}{%
\begin{tabular}{ |p{2.55cm}||p{3.5cm}|p{1.1cm}|p{1.1cm}|p{1.5cm}|p{3cm}|p{3cm}|p{2cm}|p{1cm}| }
 \hline
 \multirow{2}{1cm}{Category} & \multirow{2}{1cm}{Networks} & \multirow{2}{1cm}{Nodes} & \multirow{2}{1cm}{Edges} & \multirow{2}{1.5cm}{Num. of inputs} & \multirow{2}{3cm}{Node info} & \multirow{2}{3cm}{Edge info} & \multirow{2}{2cm}{weighting} & \multirow{2}{1cm}{Ref.} \\
 & & & & & & & & \\
 \hline
  \hline
  \multirow{6}{3cm}{Gene networks} & Stem cells 44& 44  & 547 & 4 & \multirow{6}{3cm}{Transcription factors (TF) \&  target genes (TG)} & \multirow{6}{3cm}{Molecular interaction} & unweighted & \cite{xu2013escape} \\
   & Human Tcell & 47  & 227 & 7 & & & unweighted & \cite{weinstock2024gene} \\
  & Mouse liver & 210  & 1910 & 3 & & & unweighted & \cite{fang2021grndb} \\
 & net\_p\_aeruginosa & 648 & 959 & 33 & & & unweighted & \cite{johnson2017looplessness} \\
 & net\_yeast & 662 & 1063 & 86 & & & unweighted & \cite{johnson2017looplessness} \\

 \hline
 \multirow{5}{3cm}{Power grids} & IEEE 30  & 30 & 82 & 6  & \multirow{6}{3cm}{Generators (power generation) or loads (power consumption)} & \multirow{6}{3cm}{Transmission lines} & weighted & \cite{IEEE30}  \\
 & IEEE 39 &   39  & 92 & 10 & &  & weighted & \cite{athay1979practical}\\
 &  RTS96  & 73 & 216 & 33 & &  & weighted & \cite{grigg1999ieee}\\
 & IEEE 118 &   118  & 358 & 39 & &  & weighted & \cite{118bus} \\
 & UK grid & 120  & 330   &  17 & &  & weighted & \cite{simonsen2008transient,delabays2023locating}  \\
 \hline
{Food web} & BurgessShaleS10b\_w &   48  & 243 & 6 & {Resource species (prey) \&  consumer species (predator)} & \multirow{8}{3cm}{Consumption relation} & weighted & \cite{johnson2017looplessness}\\

 \hline
 \multirow{7}{3cm}{Connectome} & C-elegans &   283  & 4690 & 86 & \multirow{7}{3cm}{Neurons or brain regions} & \multirow{7}{3cm}{Neural connections} & weighted & \cite{varshney2011structural} \\
 & Cat & 52  & 818   & 16 & & & weighted & \cite{scannell1999connectional}\\
  & Macaque 30 & 30  & 311   & 7 & & & unweighted & \cite{felleman1991distributed} \\
 & Macaque 32 & 32  & 315   & 7 & & & unweighted & \cite{felleman1991distributed} \\
 & Macaque 47 & 47  & 505   & 10 & & & unweighted & \cite{honey2007network}\\
  & Macaque 71 & 71  & 746   & 16 & & & unweighted & \cite{young1993organization} \\
 & Mouse brain & 213  & 21654   & 34 & &  & unweighted & \cite{oh2014mesoscale} \\
 \hline
 
\end{tabular}
}
    \label{tab:data}
\end{table*}

\section*{Data availability}
All data generated or analyzed during this study are included in this published article. 

\section*{Code availability}
The source code for the numerical simulations presented in the paper will be made available upon request, as the code is not required to support the main results reported in the manuscript.

\section*{Acknowledgements}
F.S. acknowledges support from grants AFOSR FA9550-24-1-0214 and Oak Ridge National Laboratory 006321-00001A. M.A. acknowledges support from the Florida State University Council on Research and Creativity SEED grant "Structure and dynamics of non-normal networks.”



%

\end{document}